\documentclass[aip,preprint]{revtex4-1}
\usepackage{graphicx} \usepackage{natbib} \usepackage{amsmath}

\draft 

\begin{document}


\title{Silicon nitride membrane resonators at millikelvin temperatures
  with quality factors exceeding $10^8$} 



\author{Mingyun Yuan}
\email{m.yuan@tudelft.nl} 
\author{Martijn A. Cohen} \author{Gary A. Steele}
\affiliation{Kavli Institute of Nanoscience, Delft University of Technology, PO Box 5046, 2600 GA, Delft, The Netherlands}


\date{\today}

\begin{abstract}
We study mechanical dissipation of the fundamental mode of
millimeter-sized, high quality-factor ($Q$) metalized silicon nitride
membranes at temperatures down to 14~mK using a three-dimensional
optomechanical cavity.  Below 200~mK, high-$Q$ modes of the membranes
show a diverging increase of $Q$ with decreasing temperature, reaching
$Q=1.27\times10^8$ at 14 mK, an order of magnitude higher than
reported before.  The ultra-low dissipation makes the membranes highly
attractive for the study of optomechanics in the quantum regime, as
well as for other applications of optomechanics such as microwave to
optical photon conversion.
\end{abstract}

\pacs{}

\maketitle 


Mechanical resonators made from silicon nitride have shown great
potential for both fundamental research and applications. They have
become platforms for studying quantum optomechanics,
\cite{Thompson2008, anetsberger2009, lahaye2004, naik2006, 
sankey2010, rocheleau2010, purdy2013a, purdy2013, eerkens2015, yuan2015} and key
elements for applications such as optical to micro- and radio-wave photon
transducers \cite{bagci2014, andrews2014} and NEMS/MEMS
sensors.\cite{Unterreithmeier2009, faust2012, fong2010} High-stress
SiN$_x$ devices typically have very high quality factors, which is a key
parameter for a mechanical resonator. In
optomechanics, low dissipation reduces the mechanical resonator's
coupling to the environment and improves the cooperativity, enabling
cooling to a lower temperature and state preparation with higher
fidelity. High quality factors also enhance the efficiency of a
transducer as well as the sensitivity of a NEMS/MEMS sensor.

Studies of the quality factor of SiN$_x$ resonators have found at room temperature $Q$
of up to $10^6$ for nanostrings,\cite{Verbridge2006}
$10^5$ for beams,\cite{Unterreithmeier2009, faust2012} $10^5$ for
trampolines, \cite{Kleckner2011} and $10^6$ for the fundamental mode
of membranes.\cite{zwickl2008} Higher modes of membranes have been
observed to have higher $Q$-factors,\cite{adiga2012, chakram2014} up
to $5\times10^7$, \cite{chakram2014} but show weaker optomechanical
coupling $g_0=\frac{\mathrm{d}\omega_0}{\mathrm{d}u}u_{zpf}$ ($\omega_0$, the cavity mode frequency; $u$, the mechanical
displacement and $u_{zpf}$, the amplitude of the zero-point
fluctuation) and have smaller mode spacing leading to a dense mode
spectrum.  Smaller membranes were previously studied down to millikelvin
temperatures, but $Q$ was relatively low due to their lower aspect
ratio.\cite{suh2013} Measurements of the $Q$ of the fundamental mode
of millimeter-sized membranes down to 300~mK demonstrated a plateau in
$Q$ below 1K at a value up to $10^7$.\cite{zwickl2008} A recent 
comprehensive review can be found in Ref.~\onlinecite{Villanueva2014}.

Here, we study the quality factor of large, high-$Q$ SiN$_x$ membranes
at temperatures down to 14~mK. We use a
three-dimensional (3D) superconducting optomechanical cavity
\cite{yuan2015} to detect the motion. Similar to previous reports, we
observe a plateau in $Q$ down to 200~mK. Below 200~mK, we observe a
new behavior of the quality factors of high-$Q$ modes that diverges
down to the lowest temperature we can measure, reaching a record of
$Q=1.27\times10^8$ for a fundamental mode at 14~mK, promising for
future applications in optomechanics in both the microwave and optical
domains.

\begin{figure}[b]
 \includegraphics[width=60mm]{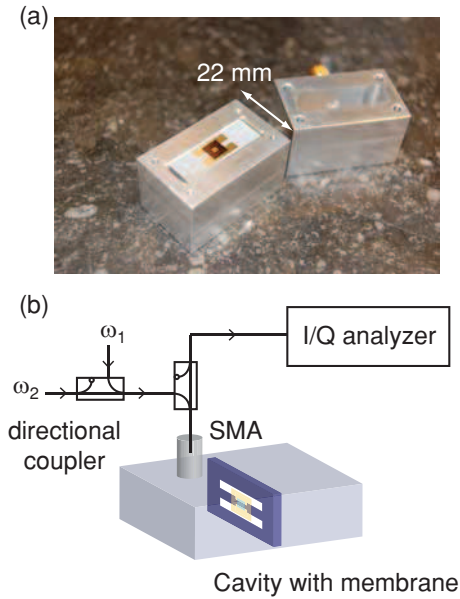}%
 \caption{\label{fig1}(a). Photograph of a physical device (Device I),
   showing two halves of the Al 3D cavity and the membrane
   resonator. (b). Schematic of the reflection measurement setup. Two
   microwave tones $\omega_1$ and $\omega_2$ are combined and launched
   into the membrane-embedded cavity placed inside a dilution
   refrigerator. The reflected signal is detected by an in-phase/quadrature (I/Q)
   analyzer. Although the schematic shows the membrane oriented vertically in the illustration, the membrane is mounted horizontally in the fridge. }%
\end{figure}

A photograph of the device is shown in Fig.~\ref{fig1}(a).  The 3D
cavity is formed by two halves of a machined Al block. An SMA
connector is partially inserted into the cavity for reflection measurement. The mechanical
resonator is a Norcada SiN$_x$ membrane which can be seen on the
sapphire support substrate. We present here results from the two
membranes we have studied at these temperatures. The membranes are
50~nm thick in a square of size $l \times l$ with $l=1.5$~mm for
Device~I and $l=1$~mm for Device~II. Device~I is stoichiometric with
$x=3/4$ and a tensile stress of 0.8~GPa. Device~II has a tensile
stress of $\leq$~0.25~GPa and $x$ is not specified by the manufacturer. We deposit a metal
layer of 20~nm of Al on the membrane which forms a capacitor with the
antenna pads deposited on the substrate.  We avoid depositing Al over
the the edges of the membrane to minimize possible mechanical losses
at the clamping points. \cite{yu2012} The motion of the membrane is
coupled to the cavity field via the antenna. The membrane is anchored
on the substrate with $\sim0.1$ $\mu$l of Bison `5 minute' 2-part
epoxy at one corner of the silicon frame. The membrane-embedded 3D
cavity is mounted with the plane of the membrane lying horizontally on
top of the antenna chip in a cryo-free dilution refrigerator (BlueFors
LD250) in which the temperature is controlled between 14~mK and
800~mK.

We measure the cavity response and mechanical motion using microwave
reflectometry as described in previous work.\cite{yuan2015} A
simplified schematic of the reflection measurement is illustrated in
Fig.\ref{fig1}(b). Microwave signals are attenuated and sent into the
cavity. The reflected signal is amplified and read out using a vector
signal analyzer (Rohde and Schwarz FSV30) which records the in- and
out-of-phase quadrature of the signal $V_i$ and $V_q$ as a function of
time within a bandwidth up to 28~MHz around a local oscillator
reference frequency. We study two membrane-cavity devices. The cavity
resonance $\omega_0$ is $2\pi\times5.23$~GHz for Device~I and
$2\pi\times5.07$~GHz for Device~II, with linewidths $\kappa$ below
300~mK of $2\pi\times56$~kHz and $2\pi\times45$~kHz, respectively. The
mechanical resonant frequency $\omega_m$ and the single-photon
coupling rate $g_0$ for the modes studied are listed in
Table~\ref{tab1}. From the resonant frequency of the fundamental modes,
we estimate the effective stress of the metalized membranes to be
0.79~GPa for Device~I and 0.09~GPa for Device~II, taking an effective
density of 3.0~g/cm$^3$. In Device II, the gap to the antenna is
3~$\mu$m. In Device I, the gap is 10~$\mu$m, resulting in a
significantly reduced $g_0$.



\begin{figure}
 \includegraphics[width=70mm]{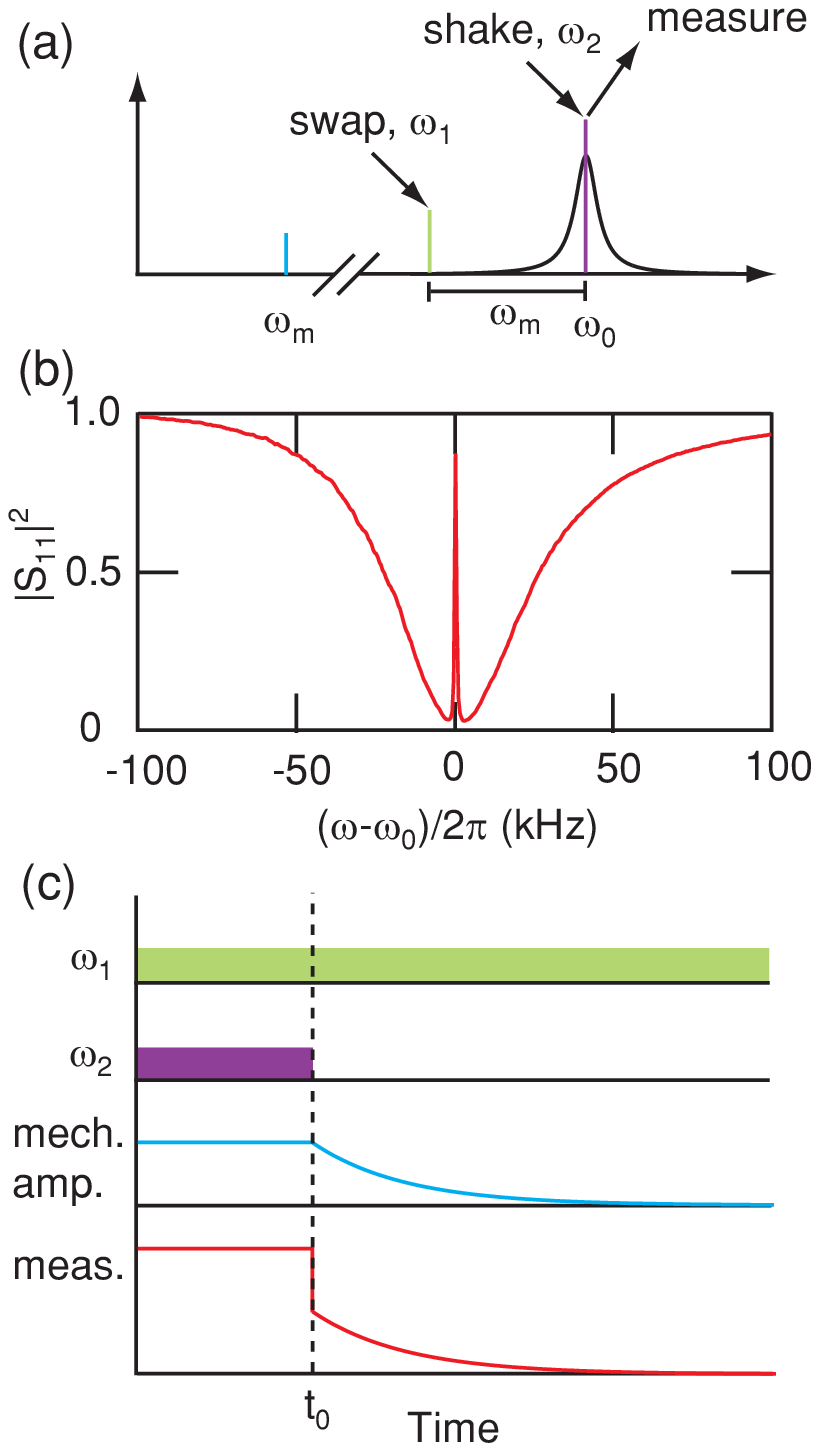}%
 \caption{\label{fig2}(a). Schematic of the measurement scheme. A swap
   tone at $\omega_1$ is used to induce an optomechanical swap
   interaction between the cavity and the membrane. Another tone at
   $\omega_2=\omega_0$ is used to shake the membrane. A vector signal
   analyzer is used to detect signal at $\omega_0$. (b). OMIT with
   $C\gg1$, showing $\vert S_{11}\vert^2$ as a function of
   frequency. To avoid optomechanical backaction, ringdown measurement
   is carried out in the $C\ll1$ regime. (c) Schematic for measurement
   of ringdown with OMIT configuration. The swap tone $\omega_1$ is
   kept on at all time while the shake tone is switched off at $t_0$,
   at which point the mechanical resonator starts to decay. The signal
   measured at $\omega_0$ is a sum of the mechanical sideband
   generated by $\omega_1$ and the reflected signal generated by
   $\omega_2$. When $\omega_2$ is switched off, the latter decays at a
   rate of $\kappa$, and the remaining mechanical component of the
   signal at $\omega_2$ then rings down on a time scale corresponding
   to $\gamma_m^{-1}$.}%
\end{figure}

To measure the quality factor, we drive the membrane at its resonance
frequency and then detect the timescale it takes for the motion to
ring down (decay). The membrane is driven optomechanically, with a
scheme based on an optomechanically-induced transparency (OMIT)
measurement. \cite{weis2010} As illustrated in Fig.~\ref{fig2}(a), two
phase-locked microwave signals are sent into the cavity: a swap tone
at $\omega_1=\omega_0-\omega_m$ and a shake tone at
$\omega_2=\omega_0$. In the presence of the swap tone and the shake
tone, there is a beating of the cavity field intensity at $\omega_m$,
giving an oscillating radiation pressure force that shakes the
drum. From the optomechanical interaction, photons at $\omega_1$ are
Raman scattered by the membrane resonator and upconverted into the
cavity resonance, producing a mechanical sideband at $\omega_0$ that
we use to measure the motion. The I/Q analyzer is set to detect the
signal at $\omega_0$ with a sample rate of 100~Hz. Fig.~\ref{fig2}(b)
shows an example of OMIT measurement taken with Device~II in the limit
of large optomechanical cooperativity
$C=\frac{4g_0^2N}{\kappa\gamma_m}\gg1$, where $N\sim750$ is the number of
photons in the cavity generated by the swap tone. To avoid backaction
of the swap tone on the motion, for the ringdown measurement we operate in a regime
where $C$ of the swap tone is sufficiently small and optomechanical damping $\gamma_{om}$ is negligible ($C\sim0.03$, $\gamma_{om}\sim0.03\gamma_m$).

\begin{table}
\caption{\label{tab1}Summary of mechanical modes studied}
\begin{tabular}{c| c| c c c}
Device&~I~&\multicolumn{3}{c}{~II}\\ \hline Mode&~(1,1)~~&~(1,1)&(1,2)&(2,1)\\ \hline
~$\omega_m/2\pi$ (kHz)~&~~242~~~&~121&193&192\\ $g_0/2\pi$
(Hz)&~~0.03~~~&~0.22&0.01&0.01\\
\end{tabular}
\end{table}

Fig.~\ref{fig2}(c) illustrates the protocol of the optomechanical
ringdown measurement.  The membrane is first driven into motion when
$\omega_1$ and $\omega_2$ are both on. The detected signal at
$\omega_0$ consists of the directly reflected signal of the shake tone at $\omega_2$
as well as the mechanical sideband generated from the swap tone at $\omega_1$. At
$t_0$ the shake tone is switched off, while the swap
tone $\omega_1$ stays on. When turning off the shake tone, the
associated microwave field at $\omega_0$ decays on a timescale corresponding to
$\kappa^{-1}$. In addition to the microwave field from the shake tone,
there is a second microwave field at $\omega_0$ that arises
from a sideband of the swap tone generated by the mechanical
motion. This second microwave field decays with a much slower
timescale corresponding to $\gamma_m^{-1}$. On the I/Q analyzer, we then
observe a signal at $\omega_0$ from shake tone $\omega_2$ that falls
off at the cavity decay rate $\kappa$ and the remaining
mechanical ringdown signal is read out and used to calculate the
mechanical amplitude as it decays.

\begin{figure}
  \includegraphics[width=70mm]{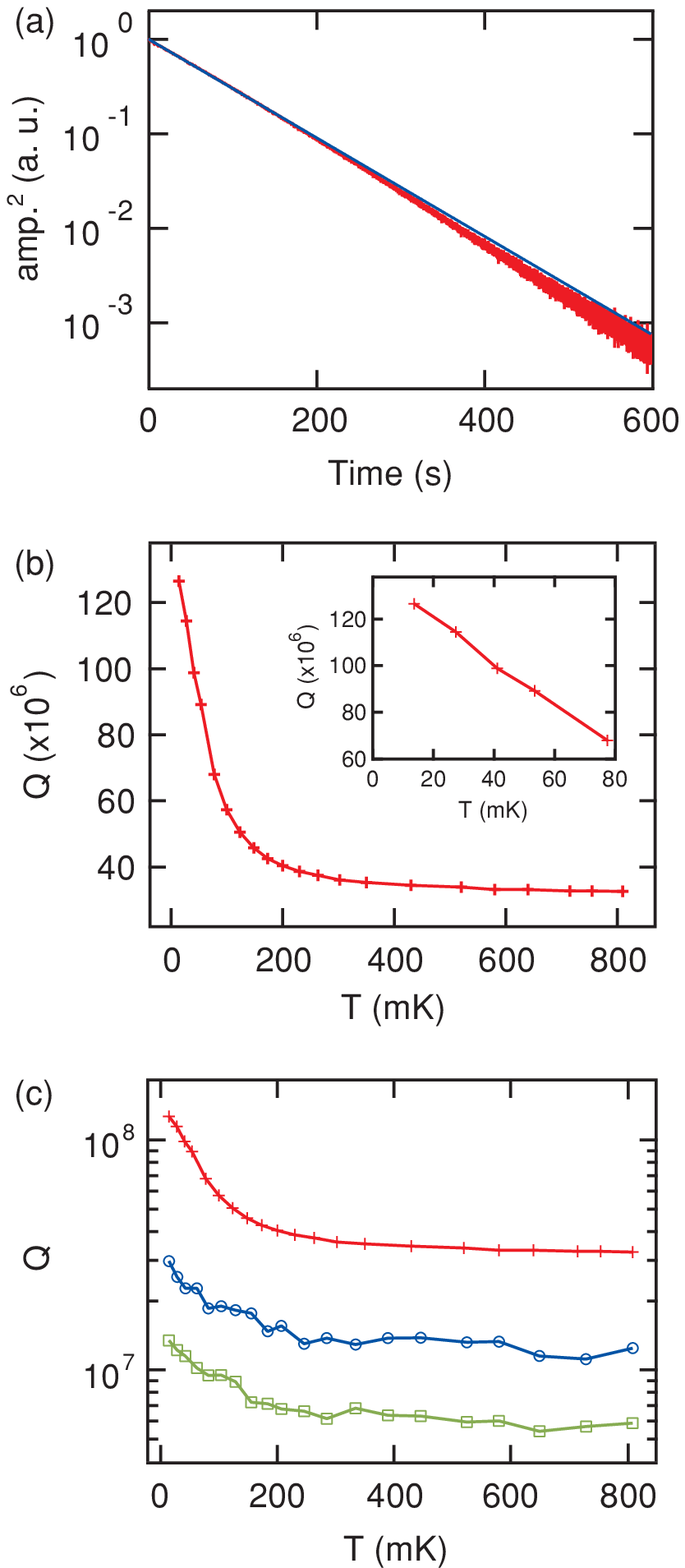}%
  \caption{\label{fig3}(a). Time trace of the mechanical amplitude
    during ringdown for Device I at 14 mK. From a fit to exponential
    decay, we extract the mechanical quality factor. Red: data; blue:
    fit. The deviation from a straight line may indicate some weak
    negative nonlinear damping. (b). Extracted quality factor $Q$ of
    the membrane resonator as a function of cryostat temperature for
    Device~I (1,1) mode. Inset: zoom-in for $T<80$~mK. The highest
    value is $Q=1.27\times10^8$ for Device I at 14~mK. (c). Extracted
    quality factor for both samples. Red: Device~I (1,1) mode; blue:
    Device~II (1,1) mode; green: Device~II (1,2) mode. Q decreases
    with increasing temperature for $14~\text{mK}<T<200~\text{mK}$,
    leveling out between 200~mK and 800~mK. }%
\end{figure}

Fig.~\ref{fig3}(a) shows an example of a ringdown trace of the
mechanical resonator taken at 14~mK.  The $y$-axis is proportional to
$V_i^2+V_q^2$, representing the square amplitude of the resonator. By
fitting the curve to an exponential decay $e^{-\gamma_m t}$,
$\gamma_m$ and the quality factor $Q=\omega_m/\gamma_m$ can be
extracted. We vary the cryostat temperature $T$ and record the
corresponding ringdown traces. The resultant $Q$-factor as a function
of temperature $T$ for Device~I (1,1) mode is plot in a linear scale
in Fig.~\ref{fig3}(b). As $T$ is decreased from 800~mK to 200~mK,
there is a relatively flat plateau in $Q$, consistent with previous
results with an optical detection scheme.\cite{zwickl2008} As $T$ is
further reduced to below 200~mK, $Q$ begins to go up, and continues
rising with no indication of saturation down to the base temperature
of $T=14$~mK. The highest value $Q=1.27\times10^8$ corresponds to
$\gamma_m=2\pi\times1.9$~mHz and a time constant for amplitude of
$\tau=2/\gamma_m=1.6\times10^2$~s.

The slight deviation from a straight line in Fig.~\ref{fig3}(a)
suggests some weak negative nonlinear damping, although the mechanical
response of the membrane is still in the linear restoring-force
regime. The amplitude unity in Fig.~\ref{fig3}(a) corresponds to
0.37~nm. For small amplitudes $<0.037$~nm, where the deviation becomes
significant, we also performed an additional exponential fit, finding
a quality factor $Q=1.16\times10^8$ for the lower amplitude regime. The Duffing critical amplitude is estimated to be 6.3~nm with the formulae in Ref.~\onlinecite{chobotov1964, kozinsky},
significantly above the excitation amplitude used here.

In Fig.~\ref{fig3}(c) we plot the $Q$ of Device~I (1,1) mode (red),
Device~II (1,1) mode (blue) and Device~II (1,2) mode (green) together
in a log scale. Quality factors of all three modes are above
$5\times10^6$ and show similar behavior, improving with decreasing
temperature below 200~mK and leveling off between 200~mK and 800~mK.
It is also interesting to note that although the mode temperature
saturates at 210~mK for Device~I and 180~mK for Device~II, $Q$
continues to go up as the cryostat is cooled down to base temperature.
The high mode temperature in the experiment is likely
related to mechanical vibration noise in the setup, which
includes only minimal vibration isolation.  The fact that we observe
an increasing $Q$ down to temperatures far below the mode temperature
suggests that the mechanism limiting the $Q$-factor is not related to
the mode occupation. A likely candidate is a physical property of the
material itself, such as surface losses.\cite{Villanueva2014} In this
case, the physical lattice could be thermalized with the fridge, while
the mechanical mode is heated out of equilibrium by the vibrational
noise.

In contrast to the other modes, Mode (2,1) of Device II has a much
lower $Q=1.2\times10^5$ that is independent of temperature from 14 to
800~mK.  Although the (1,2) mode and (2,1) are separated by only 1 kHz
in frequency, it is striking that $Q$-factors are different by orders
of magnitude.  This large difference in $Q$ could result from the
anchoring of the chip at one corner, giving very different
clamping losses through the substrate chip for the two
modes\cite{schmid2011}. With a splitting of the (1,2) and (2,1) into
modes symmetric (S) and antisymmetric (AS) with respect to the anchor
point, the AS mode could have a temperature-independent $Q$ limited by
radiation into the sapphire substrate, while the S mode would then be
limited by a different temperature-dependent mechanism.

  \begin{figure}
 \includegraphics[width=70mm]{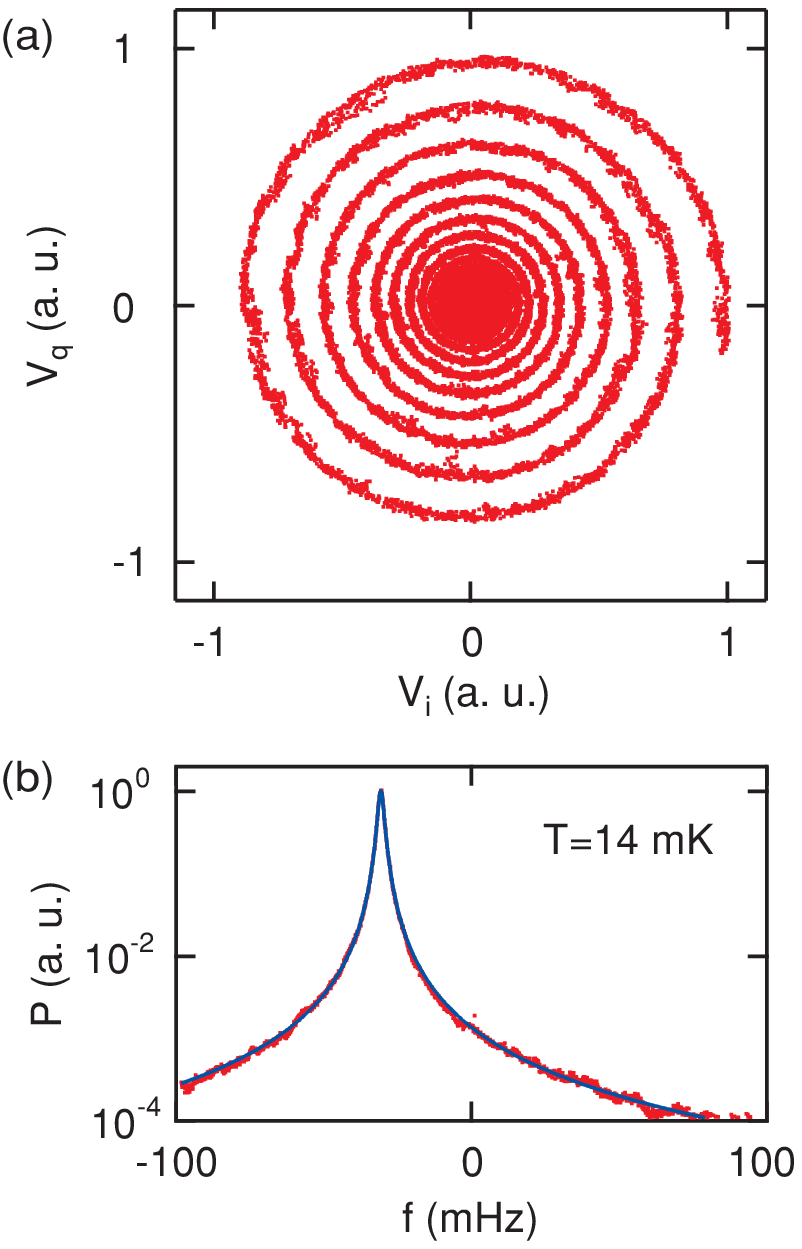}%
 \caption{\label{fig4}I/Q analysis of the ringdown trace. Swap tone is
   set to $\omega_1=\omega_0-\omega_m-\delta$,
   $\delta=2\pi\times31.9$~mHz.  (a). Out-of-phase quadrature $V_q$
   vs. in-phase quadrature $V_i$. Decay of the amplitude in combination
   with the detuning $\delta$ of the swap tone results in a spiral that
   circles around the origin at an angular frequency
   $\delta$. (b). FFT spectrum of the complex temporal I/Q trace. Red:
   data; blue: fit. From the linewidth of the resonance the
   decoherence, including dephasing and relaxation, can be
   extracted. By comparing the spectral quality factor
   $1.10\pm0.05\times10^8$ and the ringdown quality factor $1.14\times10^8$,
   the contribution of dephasing to the spectral linewidth is negligible within the error margin.}%
 \end{figure}

In the ringdown experiments performed here with the I/Q analyzer, we
are able to determine not only the energy loss rate of the mechanical
resonator, but also to characterize the dephasing of its mechanical
motion. Applying an FFT to the acquired I/Q data from sufficiently long
ringdown time trace, one can reconstruct the spectral content of the
mechanical resonance during ringdown, giving access to the spectral
linewidth $Q$-factor.\cite{schneider2014}  To do this, we slightly
detune the swap tone $\omega_1=\omega_0-\omega_m-\delta$ with
$\delta=2\pi\times31.9$~mHz and measure the I/Q data for $10^4$
seconds. In Fig.~\ref{fig4}(a) the I/Q vector plot is shown, the
$x$-axis representing the in-phase quadrature $V_i$ and the $y$-axis
the out-of-phase quadrature $V_q$. The trace forms a spiral: the
decrease of the vector length corresponds to the decay in mechanical
amplitude, and the angular frequency of the trajectory in the polar
plot is determined by $\delta$.  To reconstruct a spectrum from the
data, we perform an FFT of the complex vector $V_i+jV_q$,
$j=\sqrt{-1}$, shown in Fig.~\ref{fig4}(b). A fit to the lineshape
gives the spectral $Q = 1.10\pm0.05\times10^8$, in agreement, to within the error margin, with the ringdown quality factor $1.14\times10^8$ extracted from the same dataset, demonstrating that
the dephasing is not a significant source of decoherence for these
membrane resonators.

In conclusion, we have measured the quality factor of SiN$_x$
membranes at millikelvin temperatures with 3D optomechanical
cavities. At the base temperature of 14~mK, $Q$-factors as high as
$1.27\times10^8$ are observed for a fundamental mode, demonstrating
the exceptional performance of SiN$_x$ membranes as mechanical
resonators. This high $Q$ is achieved in the presence of an Al coating of the
membrane, expanding their potential into electrical and microwave
applications. By virtue of this low dissipation, SiN$_x$ membranes
could be an attractive test bed for quantum superposition states of massive
mechanical objects and other applications in optomechanics.



%
%

%

\begin{acknowledgments}
We thank Vibhor Singh and Simon Gr\"oblacher for
useful discussions. We acknowledge support from the Dutch Organization for Fundamental
Research on Matter (FOM) and the Netherlands Organization for Scientific
Research (NWO) through the Innovative Research Incentives Scheme (VIDI).
\end{acknowledgments}


%

\end{document}